\documentstyle{article}

\input{epsf}

\begin{document} 


\noindent
{\bf QUANTUM INTERFEROMETRY, MEASUREMENT AND
OBJECTIVITY: SOME BASIC FEATURES REVISITED}
\footnote{Presented at the 2nd INTERNATIONAL SYMPOSIUM ON FUNDAMENTAL
PROBLEMS IN QUANTUM PHYSICS, Oviedo, Spain, 21-26 July 1996 (Kluwer
Academic Publ., in press).}
\\ \\ \\ \\ \hspace*{10ex}
{\bf Markus Simonius}
\\ \\ \hspace*{10ex}
Institut f\"ur Teilchenphysik, Eidgen\"ossische Technische Hochschule,
\\ \hspace*{10ex}
CH-8093 Z\"urich, Switzerland
\\ \\ \\
The reduction paradigm of quantum interferometry and the objectivation
problem in quantum measurements are reanalyzed.
Both are shown to be amenable to straightforward
mathematical treatment within ``every-users'' simple-minded quantum
mechanics without reduction postulate etc., using only
its probabilistic content.
\\ \\ \\
{\bf 1. INTRODUCTION: POSTULATES AND INTERPRETATION}
\\ \\
Everyusers quantum mechanics, based only on it's minimal probabilistic
interpretation, is a concise and extremely successful theory without
any intrinsic contradictions. Difficulties usually arise only when
additional interpretational assumptions are introduced in order to
bridge missing links in its understanding.

It is the object of this paper to demonstrate that once the basic
axioms expressing the probabilistic content of quantum theory are
laid down problems can be solved by careful definitions and simple
but rigorous mathematics, without invoking any additional ad hoc
ingredients.

{\em Only the minimal probabilistic postulates I and II of
quantum mechanics listed below will be used. No state reduction and no
axiom of measurement etc.!\/} Postulate III is included for
completeness. It is not used but not violated either.

\begin{list}{}{\leftmargin4ex \itemsep0mm \parsep0mm} \em
\item[I  ]
States of a physical system are represented by elements $X$ of the
convex set ${\cal S}$ of density matrices on a Hilbert space ${\cal H}$
characterized by $X\ge 0$ and {\rm Tr$(X)=1$.}
\item[II ]
The probability for any given kind of event for a system in a state
represented by $X$ is expressed by an expectation value
{\rm $(A,X)\equiv$ Tr[$AX$]} where $A$ is a hermitian operator on
${\cal H}$, called an {\em observable}, which obviously must obey
$0\le A\le1$ or, equivalently, $0\le\langle\varphi|A|\varphi\rangle
\le 1$ for all $\varphi\in {\cal H},\ \|\varphi\|=1. $ The set of all
such operators is denoted by $\widehat{\cal O}$.
\item[III]
The evolution of an isolated system is reversible on ${\cal S}$
and thus unitary on ${\cal H}$.
\end{list}
Note the important fact that $(A,X)$ is convex linear on ${\cal S}$
under mixing, 
$(A,|c_1|^2X_1+|c_2|^2X_2)=|c_1|^2(A,X_1)+|c_2|^2(A,X_2)$,
as required by its interpretation as a probability and that any such
function can be represented by some $A\in\widehat{\cal O}$.

For a pure state $X=|\varphi\rangle\langle\varphi|$ with $\varphi\in
{\cal H}$, $\|\varphi\|=1$ and $(A,X)=\langle\varphi|A|\varphi\rangle$. Pure
states play no conceptually distinguished role, and in particular it
is {\em not} assumed that a system is always in some, perhaps
unknown, pure state.

In II above ``probability'' is just a number between zero and one
expressible by a formula within the theory. This number is {\em
interpreted} as the probability for the occurrence of something named
an {\em event} $e\in\{1,0\}$ where $e=1$ means that the event occurs
in a given trial and $e=0$ that it does not. There is no corresponding
formula for events, however, and actually none is needed for any
practical application. Nevertheless, it will be shown that events {\em
can} be described in quantum mechanics using only I and II above and
without violating III, where an event is {\em defined} by the
following objectivity criterion [1]:
\\
{\em The basic property defining an event and guaranteeing its
objectivity is that it can be observed or read in at least two
independent, mutually non-interfering ways with necessarily agreeing
results (zero probability for disagreement).}
\\ \\ \\
{\bf 2. SINGLE QUANTUM INTERFEROMETRY [2]}
\\ \\
Consider a typical interferometer arrangement as sketched in Fig.~1.
Its properties are well known and often discussed: even if only one
quantum (photon or neutron etc.) is inside the arrangement at a given
time, the configuration of Fig.~1a can reveal interference between the
states passing the two arms I and II provided none of them is blocked.
Correspondingly, a pure single-particle state within the
interferometer is represented by a normed wave function of the form
\begin{equation} \label{e1}
\varphi = c_1\varphi_1+c_2\varphi_2,\ \
\|\varphi\|=\|\varphi_1\|=\|\varphi_2\|=1, 
\end{equation}
where, $\varphi_1$ and $\varphi_2$ represent states passing completely
along one of the two paths I or II, respectively, in the
interferometer and have zero component in the other (and thus are
mutually orthogonal). For ideal 50:50 beam-splitting $|c_1|^2 =
|c_2|^2 = {1\over 2}$.

\begin{figure}[b]
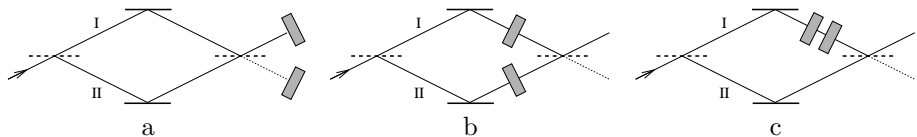

\vspace{-4mm}
\newlength{\newcm} \setlength{\newcm}{7.5mm}
\center
\begin{minipage}[c]{4.8\newcm} \center
\epsfysize=1.65\newcm
\mbox{\epsfbox{sim_figa.eps}} \\ \ a
\end{minipage}   \hspace{0.6\newcm}
\begin{minipage}[c]{4.8\newcm} \center
\epsfysize=1.65\newcm
\mbox{\epsfbox{sim_figb.eps}} \\ \ b
\end{minipage}   \hspace{0.3\newcm}
\begin{minipage}[c]{4.8\newcm} \center
\epsfysize=1.65\newcm
\mbox{\epsfbox{sim_figc.eps}} \\ \ c
\end{minipage}
\caption{Sketch of typical interferometer arrangements with
differently placed detectors.}
\label{fig1}
\end{figure}

Now consider instead coincidences between two detectors inserted into
the two paths of the interferometer as shown in Fig.\ \ref{fig1}b.
From the appearance of the wave function (\ref{e1}) one might suspect
that such coincidences should occur. But quanta (particles or photons)
are supposed to be {\em indivisible entities} and thus a {\em single
quantum} is not able to trigger both detectors in the arrangement of
Fig.\ \ref{fig1}b (nor in Fig.\ \ref{fig1}a, of course). Though all
this is much discussed textbook wisdom, one looks in vain in the
literature for a satisfying {\em derivation} of this fact which uses
only the basic structure of single particle quantum theory and does
not amount to just stating the fact as postulate in one way or
another.
But clearly it should be possible to deduce such basic
properties mathematically once the general premises of the theory are
laid down [3]. And this, in fact, turns out to be the case.

The analysis is based on the following ``silly'' theorem which is a
simple consequence of the positivity postulated in II:
\\ {\bf Theorem 1:} {\em
Let $A$ be a positive operator on ${\cal H}$ such that
$\langle\psi_1|A|\psi_1\rangle =
\langle\psi_2|A|\psi_2\rangle =0$ for given
$\psi_1,\psi_2\in{\cal H}$.  Then $\langle\psi|A|\psi\rangle = 0$ for
all superpositions $\psi = c_1\psi_1+c_2\psi_2$ between them}. \\
In order to apply this theorem to Fig.\ \ref{fig1}b with the two
detectors in coincidence one only has to remark that one of the
fundamental requirements for a coincidence between the two detectors
is that the probability for a coincidence event be zero for any state
which has zero component in one of the two arms of the interferometer,
i.e if either $c_1=0$ or $c_2$=0 in Eq.\ (\ref{e1}).  (This is
actually what careful experimenters check in order to verify that
their arrangement does not produce spurious coincidence events!)  Thus
$\langle\varphi_1|A|\varphi_1\rangle =
\langle\varphi_2|A|\varphi_2\rangle =0$ 
where $A$ is the operator describing the probability of coincidence
events. It then follows from the above theorem that
$\langle\varphi|A|\varphi\rangle = 0$ also for arbitrary $\varphi =
c_1\varphi_1+c_2\varphi_2$ and thus that {\em also for arbitrary
superpositions between the states in the two arms of the
interferometer the two detectors in Fig.\ \ref{fig1}b have zero
probability to produce a coincidence event.}

A corresponding ``reduction theorem'' is obtained similarly (using
anticoincidence between the two detectors, this time) for an
arrangement of the kind shown in Fig. 1c where now both detectors are
in the same path and it is assumed that at least the first detector
transmits (does not absorb) the quanta: If the two detectors have unit
efficiency either both of them fire or none.  Of course two-slit or
Stern-Gerlach arrangements can be analyzed correspondingly.

Thus ``what must be'' is graciously born out by the mathematical
analysis in spite of appearance of the wave function in Eq.\
(\ref{e1}), ``with nothing left to the discretion of the theoretical
physicist'' [3] (except to formulate the problems properly).
Astonishingly, in spite of its simplicity, no such treatment is found
in the literature, let alone in textbooks where it would belong.

The structure of theorem 1 and of the corresponding physical
statements is emphasized: Conclusions for arbitrary superpositions
between two states $\varphi_i$ are obtained from conditions imposed
only for the two states $\varphi_i$ themselves. There is no need to
{\em postulate} what happens in the case of superpositions.
\\ \\ \\
{\bf 3. MEASUREMENTS AND OBJECTIVE EVENTS [1]}
\\ \\
The demonstration of the emergence of objective events 
is conceptually more involved and only a rather condensed exposition
of the concepts and main result can be given here.
For a more general and detailed discussion refer to Ref. [1].

The following operational definitions of discrimination and
superpositions [1],
formulated for density operators and probabilities directly,
should be visualized with the help of the different arrangements of
Fig. 1. and related to the more familiar formulations in terms of
Hilbert space elements in case of pure states.

An observable $A\in\widehat{\cal O}$ {\em discriminates} between two
states $X_1,X_2$ if $(A,X_1)=1$ and $(A,X_2)=0$ or vice versa. Such
$A$ exists provided $X_1$ and $X_2$ are {\em orthogonal}.

A state $X$ is a {\em (general) superposition} between two orthogonal
states $X_1$ and $X_2$ with some fixed weights $|c_1|^2$ and
$|c_2|^2$ such that $|c_1|^2+|c_2|^2=1$ if 
\begin{equation} \label{superpos} \begin{array}{rl}
(A,X) =|c_2|^2 (A,X_2) & \forall A\in\widehat{\cal O} : (A,X_1)=0 \\
(A,X) =|c_1|^2 (A,X_1) & \forall A\in\widehat{\cal O} : (A,X_2)=0. 
\end{array} \end{equation}
The set of all states $X$ with this property is denoted by
${\cal S}(|c_1|^2 X_1,|c_2|^2 X_2)$.
It is obviously convex and contains the incoherent mixture 
$|c_1|^2 X_1+|c_2|^2 X_2$.

An observable $A$ is {\em sensitive to
interference} between $X_1$ and $X_2$ if 
$(A,X) \ne (A,|c_1|^2X_1+|c_2|^2X_2)$ for some
$X \in{\cal S}(|c_1|^2X_1,|c_2|^2 X_2)$.

An immediate consequence of this definition of superpositions is the
following generalization of theorem 1 (and thus its
implications!) to non-pure states:\\
\noindent{\bf Theorem $1'$:} {\em
If in an arrangement $(A,X_1)=(A,X_2)=0$ for two
orthogonal states $X_1,X_2$, then $(A,X)=0$ also for any
superposition $X$ between these two states.}

A central role plays the careful analysis and definition of the
concept of measurement:
In a measurement information on the state $X$ of an object is
distributed onto different, at least two, separated channels
(identified by greek upper indices) consisting of different systems on
which mutually non-disturbing observations, using channel observables
or {\em readings} $A^\mu\in{\cal O}^\mu,\ \mu=1, 2,...$, may be
performed.

A given measurement is characterized by a function $m(\{A^\mu,\ \mu\in
M\};X):\ \otimes_{\mu\in M}{\cal O}^\mu\otimes{\cal S} \to [0,1]$
where $M$ is the subset of channels which are read. It will be denoted
also $m(A^\mu;X)$ in case only one channel $\mu$ is read or
$m(A^\mu,A^\nu;X)$ in case two channels $\mu$ and $\nu$ are read
simultaneously etc.. It is to be interpreted as the probability for
{\em coincidence events} $\prod e^\mu,\ \mu\in M$, i.e
$m(\{A^\mu,\mu\in M\};X)=Prob(\prod_{\mu\in M} e^\mu=1) $ where
$e^\mu$ is the event obtained by the reading $A^\mu$.

The function $m$ must be (convex) linear in ${\cal S}$ and all ${\cal
O}^\mu$ in analogy to the corresponding properties of $(A,X)$
appropriate for probabilities as discussed.  In addition it has to
fulfill the following important requirement expressing {\em
separability or mutual non-disturbance} of readings of different
channels:
\begin{equation} \label{sep} 
m(A^\mu,A^\nu;X)+m(A^\mu,\bar{A^\nu};X)=m(A^\mu;X)
\end{equation} 
($\mu\ne\nu$) for all $A^{\mu,\nu}\in{\cal O}^{\mu,\nu}$ independent
of $A^\nu$, and correspondingly for an 
arbitrary number of channels.
Here the {\em complement} $\bar{A}$ of an observable $A$ is defined by
$(\bar{A},X)=1-(A,X)$ for arbitrary state $X$ which can be thought of as
being the observable obtained from $A$ by the replacement of the
corresponding  event $e$ by its negation $\bar{e}=1-e$ (i.~e. by
inverting the output).
The l.s.h of Eq. \ref{sep} then means that the event
corresponding to $A^\nu$ is 
ignored though it has been obtained, and the r.h.s that no observation
of channel $\nu$ is performed at all.

A channel $\mu$ of a measurement {\em discriminates} between
$X_1$ and $X_2$ if reading of that channel alone, without further
observation, allows one to discriminate between $X_1$ and $X_2$. A
corresponding reading $A^\mu$ {\em discriminates $X_1$ against $X_2$}
if $m(A^\mu;X_1)=1$ and $m(A^\mu;X_2)=0$. This defines what a
measurement is good for.

The fact that for arbitrary fixed readings $A^\mu,\
\mu\in M$ a measurement realizes some observable $A\in\widehat{\cal O}$
such that $(A,X)=m(\{A^\mu,\ \mu\in M\};X)$ together with the defining
Eq. (\ref{superpos}) of superpositions then leads to the central result:
\\ 
{\bf Theorem 2 (Probability and Objectivity):} {\em Consider a
measurement with readings $A^\mu,A^\nu$ for two different channels
which both discriminate $X_1$ against $X_2$ and assume
that $X\in{\cal S}(|c_1|^2X_1,|c_2|^2X_2)$ is any superposition
between the $X_i$. Then
\begin{equation} \label{prob}
m(A^\mu,A^\nu;X)=m(A^\mu;X)=m(A^\nu;X)=\|c_1\|^2
\end{equation}
and
\begin{equation} \label{obj} 
m(A^\mu,\bar{A^\nu};X)=m(\bar{A^\mu},A^\nu;X)=0 
\end{equation}
implying that for all $X\in{\cal S}(|c_1|^2X_1,|c_2|^2X_2)$ two such
discriminating readings performed on separate channels have zero
probability to disagree.} The last statement is exactly the defining
property of an objective event as given in section 1 which leads to
the following\\ {\bf Corollary (Objective Events):} {\em A measurement
which discriminates between two states $X_1$ and $X_2$ according to
the condition of theorem 2, performed on any superposition $X$ between
them, leads for each trial to an (objective) event for an observable
$A$ which discriminates between $X_1$ and $X_2$.}\\ The {\em value} of
the event is not predicted in general, only its probability according
to Eq. (\ref{prob}). However, under the condition of theorem 2, the
probabilistic element is completely eliminated in the correlation
between two discriminating channels. Reading one of them one can or
could with certainty predict the outcome of an eventual reading of the
other: ``The moon is {\em objectively} there even if nobody looks.''

The mutual independence of the readings of different channels required
by the definition of objectivity is contained in the important
separability condition (\ref{sep}). Besides this, the central
ingredient of the proof of Eq. (\ref{obj}) is (again) theorem $1'$:
The condition of theorem 2 together with the separability condition
(\ref{sep}) which yields $m(A^\mu,\bar{A^\nu};X)\le m(A^\mu;X)$ etc.
implies $m(A^\mu,\bar{A^\nu};X_i) = m(\bar{A^\mu},A^\nu;X_i) = 0$
which then must hold for any superposition $X$ between $X_1$ and $X_2$
by theorem $1'$.

Thus objective events emerge automatically from the definition of
measurements. It is emphasized that {\em only the mathematical
properties of the functions $m(\cdot\, ;\cdot)$ and $(\cdot\, ,\cdot)$ are
actually used in the proof without reference to their interpretation
in terms of events. Moreover, only the response of each channel
separately to the two states $X_i$ is presupposed and, as in section
2,  nothing about
superpositions between them nor about correlations among different
channels.}

It can be shown [1] that the definition and emergence of objective
events does not violate any principle of quantum mechanics. The
rational behind this is quite simple: {\em One just cannot violate the
defining equations (\ref{superpos}) of superpositions however
complicated one chooses the arrangement for an observation to be.}

{\em The crucial feature for objectivity and consistency turns out to
be the distribution of the information onto different channels.}
In the classical domain this happens continuously due to
the abundance of channels consisting of light. 
\\ \\ \\
{\bf REFERENCES}
\\ \\
1. M.~Simonius, {\it Helv.~Phys.~Acta\/} {\bf 66}, 721 (1993); {\bf
65}, 884 (1992).
\\
2. M.~Simonius, {\it Helv.~Phys.~Acta\/} {\bf 69} Separanda, 15
(1996). 
\\
3. J.~S.~Bell, {\it Physics World\/} {\bf 3} No. 8, August 1990, p.
33.

\end{document}